\documentclass[runningheads]{llncs}

\usepackage[T1]{fontenc}
\usepackage[pdftex]{graphicx}
\usepackage{amsmath,amssymb}
\usepackage{tablefootnote}
\usepackage{hyperref}[draft]
\usepackage{url}
\usepackage{tabularx}
\usepackage{comment}
\usepackage{multirow}
\usepackage{enumitem}
\usepackage{xspace} 
\newcommand{\methodName}{RAPID\xspace}
\usepackage{soul}

\begin{document}
\title{\methodName{}: Robust APT Detection and Investigation Using Context-Aware Deep Learning}
\titlerunning{Robust APT Investigation Using Deep learning}

\author{Yonatan Amaru \and Prasanna Wudali  \and Yuval Elovici \and Asaf Shabtai}
\institute{Ben-Gurion University of the Negev, Israel \\
\email{amaruy@post.bgu.ac.il, \{wudalipr,elovici,shabtaia\}@bgu.ac.il}}

\authorrunning{Amaru et al.}

\maketitle

\begin{abstract}
Advanced persistent threats (APTs) pose significant challenges for organizations, leading to data breaches, financial losses, and reputational damage. 
Existing provenance-based approaches for APT detection often struggle with high false positive rates, a lack of interpretability, and an inability to adapt to evolving system behavior. 
We introduce \methodName{}, a novel deep learning-based method for robust APT detection and investigation, leveraging context-aware anomaly detection and alert tracing. 
By utilizing self-supervised sequence learning and iteratively learned embeddings, our approach effectively adapts to dynamic system behavior. 
The use of provenance tracing both enriches the alerts and enhances the detection capabilities of our approach.
Our extensive evaluation demonstrates \methodName{}'s effectiveness and computational efficiency in real-world scenarios. 
In addition, \methodName{} achieves higher precision and recall than state-of-the-art methods, significantly reducing false positives.
\methodName{} integrates contextual information and facilitates a smooth transition from detection to investigation, providing security teams with detailed insights to efficiently address APT threats.
\end{abstract}

\section{Introduction \label{sec:introduction}}
Advanced persistent threats (APTs) are highly sophisticated and targeted cyberattacks that pose significant risks to organizations, leading to data breaches, intellectual property theft, financial loss, and reputational damage. 
Recent high-profile incidents, such as the SolarWinds supply chain attack~\cite{lazarovitz2021deconstructing} and the Microsoft Exchange Server vulnerability exploitation~\cite{pitney2022systematic}, underscore the critical need for robust APT detection and investigation capabilities that can adapt to the evolving nature of these threats.

Provenance data analysis has emerged as a promising approach for APT detection and investigation. 
Rule-based methods for provenance data analysis~\cite{hossain2017sleuth,milajerdi2019holmes,hassan2020tactical,kurniawan2022krystal} can detect known attack patterns but are impractical due to the need for expert labeling and their inability to detect zero-day attacks. 
To address these limitations, many studies have investigated anomaly-based methods~\cite{alsaheel2021atlas,yu2019needle,hassan2019nodoze,wang2020you,hossain2020combating,du2017deeplog,zhang2019robust,devlin2018bert,liu2019log2vec,han2020unicorn,wang2022threatrace,zengy2022shadewatcher,jia2023magic,yang2023prographer}. 
These methods, which focus on learning normal system behavior and detecting deviations, are capable of identifying novel attack vectors without the need for attack data, which may not always be available.

However, in the practical application of anomaly-based methods alert fatigue remains a primary challenge, as it can be difficult to maintain an acceptable balance between high recall and a low false positive rate due to their sensitivity to normal variability (natural noise) in system behavior.
While several recent studies~\cite{alsaheel2021atlas,wang2022threatrace,zengy2022shadewatcher,jia2023magic,han2020unicorn,Li_2024} proposing anomaly-based methods for APT detection have reported good results, the proposed methods have two main limitations: (1) they rely on large training datasets and disproportionally small testing windows, which does not align with the dynamic and continuously evolving nature of real-world data environments, and (2) they suffer from data leakage~\cite{wang2022threatrace,jia2023magic}, since the models were inadvertently trained on future data, which may result in exaggerated detection performance. 
Thus, although these methods produced good results, further research is needed to address their limitations and improve their effectiveness and reliability; addressing these issues will increase their adoption and real-world application.

In light of the above-mentioned research gaps and limitations, we propose \methodName{}, a robust anomaly-based APT detection and investigation framework that leverages context-aware deep learning on provenance data to reduce alert fatigue and ensure high recall in evolving system environments. 
\methodName{} combines self-supervised learning and provenance analysis to account for the dynamic nature of evolving systems, offering a practical and effective solution for real-world application. 
This integration enables \methodName{} to detect unknown attacks with very low false alarms under realistic conditions (i.e., ensuring no data leakage), thus addressing critical gaps in existing APT detection methods.

\methodName{} consists of two phases: detection and tracing. 
In the detection phase, \methodName{} uses object embeddings to enrich the anomaly detector with contextual information. 
By learning the object embeddings as dense vector representations of system entities based on their interactions, \methodName{} captures the nuanced relationships between processes, files, and network connections. 
These embeddings can be iteratively adjusted during inference to adapt to changes in system behavior, enabling \methodName{} to effectively distinguish between benign and malicious activities in evolving environments. 
This context-aware approach helps reduce false positives and maintain high recall, addressing the limitations of traditional anomaly-based methods that struggle to adapt to changing system dynamics.

In the tracing phase, the challenge of generating clear and actionable attack narratives is addressed by grounding the alerts in the provenance data, further reducing false positives. 
Unlike existing methods that rely on predefined rules or attempt to trace attacks from a single entry point, \methodName{} leverages the anomalies detected in the first phase, using them as multiple starting points for the reconstruction of potential attacks. 
By intelligently back-tracking and forward-tracing the provenance data and filtering the most relevant events, \methodName{} reconstructs the complete attack narrative, providing security analysts with a comprehensive view of the attack kill chain. 
This approach minimizes alert fatigue by presenting a coherent and precise attack story, enabling analysts to quickly understand and respond to threats.

To validate \methodName{}'s effectiveness in real-world settings, we conducted an extensive evaluation using three diverse and well-established datasets: the CADETS, THEIA~\cite{darpa}, and Public Arena~\cite{publicarena} datasets. 
Unlike previous studies that introduced data leakage and used unrealistic train-test splits, we employ a evaluation strategy that closely emulates real-world scenarios and mitigates data leakage. 
 
Our experiments show that \methodName{} consistently outperforms state-of-the-art methods across multiple granularity levels, achieving near-perfect detection accuracy, significantly reducing false positives, and generating precise alerts that facilitate efficient incident response and investigation.

\noindent The main contributions of this paper are:

\begin{itemize}[label=$\bullet$]

\item \textbf{\methodName{}}: A novel context-aware deep learning framework for APT detection and investigation that reduces alert fatigue while maintaining high recall in evolving system environments.

\item \textbf{Dual-purpose object embedding technique}: A novel strategy that utilizes iteratively adjustable object embeddings for enhanced accuracy and complete attack narrative reconstruction.

\item \textbf{Comprehensive evaluation}: A comprehensive evaluation demonstrating \methodName{}'s superior performance compared to state-of-the-art methods using appropriate train-test splits and small training windows.

\item \textbf{Real-world applicability}: Ensuring \methodName{}'s suitability for large-scale system logs in enterprise networks, making it a practical solution for industry adoption.

\item \textbf{Open-source implementation}: Fostering further research and adoption by making \methodName{}'s implementation publicly available.

\end{itemize}

\section{Background}\label{sec:background}

The increasing sophistication of APTs necessitates the development of robust and effective detection methods. 
Provenance graphs have become an important tool in this domain~\cite{zipperle2022provenance,inam2023sok}. 
This section begins with a motivating example of a provenance graph, which is followed by a description of the assumed threat model, and concludes with a review of prior research, highlighting its limitations. 

\subsection{Motivating Example} 

\begin{figure}
    \centering
    \includegraphics[width=\textwidth]{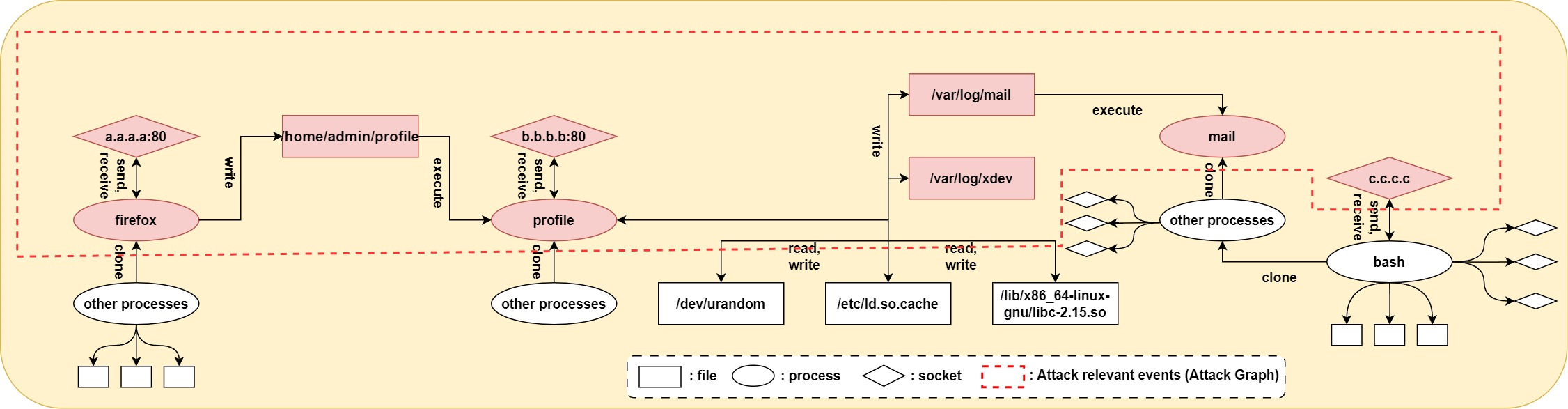}
    \caption{An example of a provenance graph containing host-based intrusion behavior from the DARA eng.3 THEIA dataset.}
    \label{fig:example}
\end{figure}

Provenance graphs are a common approach for representing system audit data~\cite{zipperle2022provenance,inam2023sok}. 
In these graphs, subjects (i.e., processes) and objects (i.e., processes, files, and sockets) are represented as nodes, and links represent the interactions between the subjects and objects.

Figure~\ref{fig:example} presents the provenance graph of an APT attack scenario from the DARPA eng.3 THEIA dataset~\cite{darpa}. 
The attack begins when a user visits a malicious website (\textit{a.a.a.a:80}) that exploits a backdoor in the \textit{Firefox 54.0.1} process to download a file (\textit{/home/admin/profile}). 
The compromised Firefox process facilitates a data transfer between \textit{a.a.a.a:80} and \textit{/home/admin/profile}. 
After a brief pause, the downloaded file is executed with root privileges, connecting to a command and control server (\textit{b.b.b.b:80}) and implanting another file (\textit{/var/log/mail}). 
This implanted file's privileges are then elevated and the file is executed. 
The executed file connects to \textit{c.c.c.c} to carry out the attacker's intended activities. The attack activity (outlined in red) is recorded, along with benign background activity, in the provenance graph. 

As new events are captured by the auditing system, provenance graphs can become extremely large and difficult to manage.
Therefore, provenance-based threat detection systems aim to extract precise attack sub-graphs from immense provenance graphs by filtering out benign background noise (i.e., spurious dependencies).

\subsection{Threat Model}

Our threat model focuses on external APT actors, excluding insider threats, supply chain attacks, and physical access to targeted systems. 
Within this scope, the proposed framework performs a comprehensive analysis of the entire attack chain, from initial compromise to lateral movement and data exfiltration.

We assume that attackers' actions leave discernible traces in system logs, enabling detection and tracing, although attacks may span extended periods of time. 
We consider the auditing system to be uncompromised, serving as a trusted computing base (TCB), and tamper-proof, ensuring the integrity and reliability of the collected system logs~\cite{bates2015trustworthy}. 
Furthermore, we assume that the provenance graphs constructed from these logs capture all relevant system entities and their interactions with sufficient granularity for effective APT detection and tracing, done in previous studies~\cite{sharma2023advanced,ghafir2018detection,inam2023sok,hossain2017sleuth,milajerdi2019holmes,hassan2020tactical}.

\subsection{Prior Research}

Provenance-based methods for APT detection were developed to counter increasingly complex cybersecurity threats. 
Rule-based techniques~\cite{milajerdi2019holmes,yu2019needle} employ predefined security policies and heuristic rules to pinpoint attack patterns. 
While these methods provide detailed fine-grained event-level detection~\cite{inam2023sok} and may have low false positive rates, they require substantial manual intervention and struggle to detect zero-day exploits~\cite{hossain2020combating}.

Anomaly-based methods, including statistical approaches~\cite{hossain2017sleuth,hassan2019nodoze,wang2020you,kurniawan2022krystal,dong2023distdet,Li_2024}, path-based methods~\cite{du2017deeplog,zhang2019robust,guo2021logbert,alsaheel2021atlas}, and graph-based methods~\cite{han2020unicorn,wang2022threatrace,zengy2022shadewatcher,jia2023magic,yang2023prographer}, have shown promising results in detecting APTs. 
However, these methods often fail to adapt to evolving system behavior over time, leading to high false positive rates. 
Moreover, their alerts lack interpretability, making it difficult for security analysts to understand and investigate detected anomalies.

Despite the promising results reported in recent important studies~\cite{alsaheel2021atlas,wang2022threatrace,zengy2022shadewatcher,jia2023magic,han2020unicorn,Li_2024}, the approaches used in these studies have two main methodological limitations. 
First, they depend on large training datasets and disproportionally small testing windows, which does not align with the dynamic and continuously evolving nature of real-world data environments. 
Second, in some studies, data leakage was identified~\cite{wang2022threatrace,jia2023magic}, where models were inadvertently trained on future data, which may result in exaggerated detection performance.
These limitations highlight the need for further research to refine the efficacy and reliability of these methods, ensuring their broader adoption and practical application in industry settings~\cite{van2019sok,dong2023we}.

To conclude, the main limitations identified in the proposed methods and prior studies are: 
(1) rule-based methods require manual maintenance and are unable to detect zero-day attacks;
(2) anomaly-based methods fail to adapt to evolving system behavior over time, resulting in high false positive rates; 
(3) alerts generated by anomaly-based methods lack interpretability, making it difficult for security analysts to investigate detected anomalies;
(4) in many cases, the evaluation performed used training data that do not reflect real-world scenarios, leading to unrealistic (incorrect) performance results; and
(5) existing methods face challenges with scalability and have high computational demands, limiting their practical deployment.

To address these limitations and provide an effective APT detection and investigation solution, we propose \methodName{}. 
\methodName{} employs self-supervised sequence-based learning and iteratively updated embeddings to effectively capture dynamic system behavior. 
Our framework's detection capabilities are enhanced with the use of provenance tracing to enrich alerts, providing more context and consequently reducing false positives. 
Additionally, \methodName{} utilizes unsupervised embedding techniques to adapt to evolving systems and focus its tracing efforts. 
Designed to be computationally efficient, \methodName{} is suitable for deployment in real-world settings. 
\section{Proposed Method: \methodName{} \label{sec:method}}
\subsection{Overview}
\methodName{} is a novel framework for APT detection and investigation that combines anomaly detection, provenance graph analysis, and self-supervised learning techniques to generate informative, context-rich alerts. 

Its high performance, which is demonstrated in our evaluation, stems from the fact that it focuses on self-supervision, the data transfer layer, and the generation of informative alerts with clear attack narratives, while considering space complexity to ensure scalability.

\begin{figure*}[htbp]
    \centering
    \includegraphics[width=1.0\textwidth]{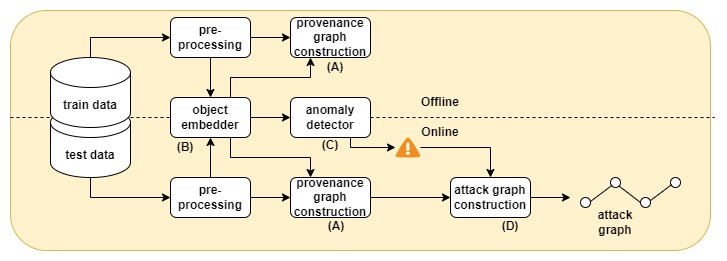}
    \caption{\methodName{}'s workflow.}
    \label{fig:MethodSchema}
\end{figure*}

As can be seen in Figure~\ref{fig:MethodSchema}, \methodName{}'s workflow starts with the construction of a provenance graph from system logs (A), which captures the relationships between system entities. 
Simultaneously, \methodName{} learns object embeddings (B) to represent the behavioral characteristics of these entities. 
The embeddings are then used by a Bi-LSTM-based anomaly detector to identify anomalies (C). 
Finally, \methodName{} grounds the anomalies in the provenance graph, using clustering and kill chain analysis to outline potential attack narratives (D). 
This process results in informative and precise alerts that enable security analysts to effectively investigate potential attacks.

\subsection{Data Collection and Processing}
\methodName{} leverages data from established auditing frameworks, such as the Linux Audit Subsystem (Auditd), Event Tracing for Windows (ETW), and FreeBSD's DTrace, which perform system call interception to capture comprehensive provenance data~\cite{bates2015trustworthy}. The event logs are collected, abstracted at the system call level, and standardized into a uniform structure: \textit{<uuid, timestamp, process UUID, process name, event, object type, object data, object UUID>}. This structure enables provenance-based analysis by characterizing each event according to its subject, object, action, and timestamp. \methodName{} focuses on events related to data transfers between system entities, which are essential for mapping information flow~\cite{barre2019mining}.

\subsection{Provenance Graph Construction\label{sec:graph_construction}}

\methodName{} constructs a provenance graph to capture a system's information flow by abstracting system events to the data transfer layer~\cite{barre2019mining,hassan2019nodoze}. To refine the graph and reduce potential false positives, \methodName{} employs two key techniques:
\begin{enumerate}
    \item Filtering out noise processes and files that do not actively participate in the data flow, based on the principle that an entity must engage in data transfer to be considered part of an attack chain~\cite{hossain2017sleuth,hassan2020omegalog}.
    \item Propagating from external entry points~\cite{hossain2017sleuth}, under the assumption that an attack must originate from an external source.
\end{enumerate}

\noindent Propagating from the external entry points into the graphs ensure capturing only the potential paths for attack propagation and thus eliminating benign nodes and events form introducing spurious dependencies. 
The resulting provenance graph, with nodes representing processes and objects and edges signifying the data transfer direction, serves as the foundation of \methodName{}'s correlation techniques, which are used to identify malicious activities.

By focusing on data transfer events and applying filtering techniques, \methodName{} reduces the space complexity of the provenance graph, making it more practical for real-world deployment. The abstraction of system events to the data transfer layer allows \methodName{} to capture the essential information flow while minimizing the storage and processing overhead associated with large-scale provenance data.

\subsection{Object Embedding\label{sec:embedding}}

\begin{figure}[htbp]
\centering
\includegraphics[width=1.0\textwidth]{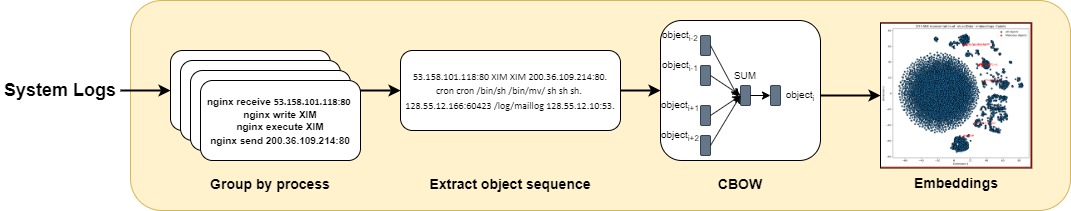}
\caption{\methodName{}'s object embedding.}
\label{fig:embedding_schema}
\end{figure}

\methodName{} leverages object embeddings to capture the behavioral characteristics of system entities, which are employed for both anomaly detection and attack tracing in the provenance graph. 
The embedding process consists of grouping the data by process, extracting object sequences, and learning embeddings using the continuous bag-of-words (CBOW) model, as illustrated in Figure~\ref{fig:embedding_schema}.

\noindent \textbf{Embedding Model.}
The data is grouped by process, with the event sequence of each process forming a "sentence" where nodes (files, ports, processes) function as "words." 
The CBOW model, a variant of Word2Vec~\cite{mikolov2013distributed,rehurek_lrec}, is used to learn object embeddings by predicting the target word based on its surrounding context. 

\noindent \textbf{Dynamic Adaptation.}
To address the challenge of out-of-vocabulary (OOV) objects and evolving system behavior, \methodName{} employs iteratively adjusted embedding construction. 
The embedding model is initially trained on a baseline vocabulary. 
As the system logs are updated, \methodName{} performs batch training to adjust and expand the embeddings, ensuring that they accurately reflect the system's evolving behavior.

The object embeddings serve a dual purpose in \methodName{}, enhancing both anomaly detection and attack tracing capabilities by capturing the behavioral characteristics of system entities. 
This simple yet effective method of obtaining them is aimed at reducing computational overhead during inference.

\subsection{Anomaly Detection\label{sec:anomaly_detection}}

\methodName{} employs a bidirectional RNN with LSTM for threat detection,
which is a lightweight alternative to complex models like transformers. 
This approach balances the tradeoff between the need to learn benign system activity with minimal training
data and the risk of over-generalizing and misclassifying attacks as normal
behavior~\cite{zhang2019robust}. 
By integrating object embeddings
(Section~\ref{sec:embedding}), \methodName{} enhances the input data's fidelity,
thereby improving detection accuracy.

The anomaly detection logic capitalizes on the model's ability to accurately predict masked events in a sequence, flagging deviations from predicted patterns as potential anomalies while preserving critical event context for subsequent alert tracing~\cite{devlin2018bert}.

\begin{figure*}[htbp]
\centering
\includegraphics[width=1.0\textwidth]{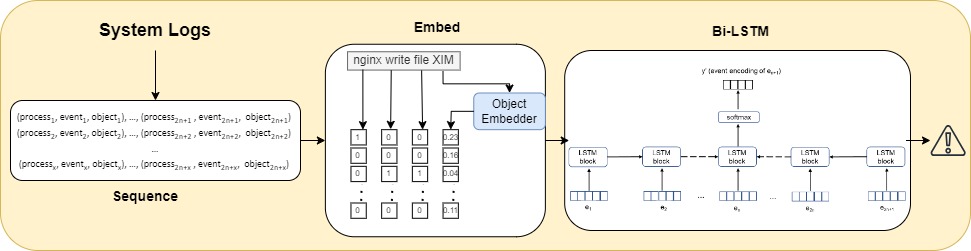}
\caption{Anomaly detection model.}
\label{fig:ADSchema}
\end{figure*}

\noindent \textbf{Neural Network Architecture.}
The neural network architecture, depicted in Figure~\ref{fig:ADSchema}, consists of an embedding layer that concatenates the contextual object embeddings with the event and process embeddings. 
The concatenated embeddings are then fed to the model, which consists of Bi-LSTM layers, two linear layers, and a softmax activation layer, for event-type classification.

Sequences are constructed based on the chronological order of events, and for each event process entity, the event type, object type, and object entity are extracted. 
A sliding window of \textit{N} events is used to form sequences, in which the central event is masked.

\noindent \textbf{Training.}
The anomaly detection model is trained on attack-free system logs to model benign behavior, using weighted learning to handle event-type imbalances. 
Cross-entropy loss and an Adam optimizer with a decaying learning rate are employed for training, which is stopped when the loss stabilizes in order to prevent overfitting.\\

\noindent \textbf{Anomaly Score Threshold Setting.\label{sec:threshold}}
Anomaly scores are assigned to each event, calculated as $anomaly(event_i) = 1 - predicted_i$, where $predicted_i$ represents the predicted softmax probability of the true event type for event $i$. 
The empirical cumulative distribution function (ECDF) with $\alpha=1$ is used to determine the threshold for anomalous events, based on the anomaly score distribution of the validation set~\cite{du2017deeplog}. 
This method ensures that the threshold is set based on the statistical properties of the anomaly score distribution, adapting to the specific characteristics of the model and data.

\subsection{Alert Tracing}\label{sec:alert_generation}
Following the anomaly detection phase, in \methodName{}'s tracing phase, the detected anomalies are integrated with the provenance graph to derive detailed and actionable
alert narratives. 
By leveraging both the anomaly scores and object embeddings, \methodName{} ensures
that in the transition between detection and tracing all critical information is retained,
thereby enabling the robust analysis of potential attack kill chains.

\begin{figure}[htbp]
    \centering
    \includegraphics[width=0.9\textwidth]{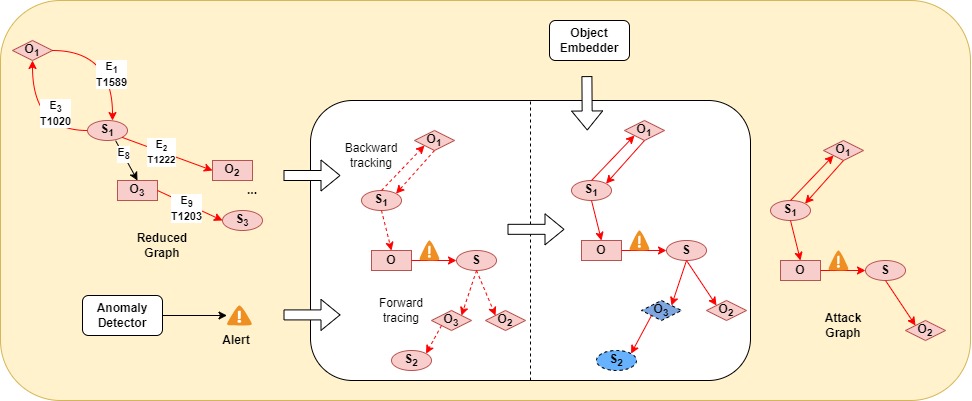}
    \caption{Illustration of the alert tracing process in which anomalies are integrated into the provenance graph.}
    \label{fig:alert_generation}
\end{figure}

\noindent The process, as illustrated in Figure~\ref{fig:alert_generation}, involves several steps designed to maintain data fidelity and enhance interpretability:
\begin{enumerate}
    \item \textbf{Anomaly Provenance Propagation:} Detected anomalies are mapped back to the provenance graph, identifying the corresponding nodes and edges. \methodName{} performs forward and backward propagation from each anomalous edge in order to extract all behavior associated with the anomaly, capturing the entire context surrounding the anomalous activity.
    \item \textbf{Entity Clustering and Filtering:} \methodName{} leverages the contextual object embeddings to cluster objects using DBSCAN~\cite{schubert2017dbscan}. The optimal number of clusters is determined using the nearest neighbors and elbow methods in a unsupervised manner. 
    The propagated sub-graph is then filtered, retaining only the objects that are in the same cluster as the anomalous nodes. 
    In this step, the main activity the anomaly is involved in is identified, allowing \methodName{} to focus on the most relevant components of the attack chain, while reducing the sub-graph's size and improving computational efficiency.
    \item \textbf{Alert Merging:} To further reduce the number of alerts and provide a more concise representation of the attack, \methodName{} merges sub-graphs with overlapping anomalies. The merged graphs, referred to as attack graphs, present potentially malicious activity and serve as the foundation for subsequent alert prioritization and investigation.
\end{enumerate}

\noindent \methodName{}'s alert tracing approach leverages unsupervised learning and graph-based analysis to robustly trace potential attacks through the provenance graph, providing a comprehensive and accurate representation of the attack kill chain without requiring labeled data or extensive manual configuration.

\section{Evaluation Setup\label{sec:experiment}}
In this section, we provide a detailed description of the experimental setup and method used to obtain the ground truth. \methodName{}'s performance is evaluated on three diverse public datasets from common auditing systems. We compare its performance to that of state-of-the-art methods, analyze the quality and traceability of the alerts generated by \methodName{}, assess the impact of the object embeddings, and examine its scalability and performance overhead.

\subsection{\methodName{} Implementation}

We implemented \methodName{} in Python 3.9, using custom parsers for each data source to extract relevant logs and filter out incomplete records. The object embedding model was constructed using Gensim~\cite{rehurek_lrec}, with an embedding size of 100. The anomaly detection model was implemented using PyTorch~\cite{torch}, with a neural network architecture consisting of four bidirectional LSTM layers, each with 256 neurons. A window length of 21 events was used. Training was performed with a batch size of 4,096, a learning rate of 1e-3, and a weight decay rate of 1e-5. The anomaly threshold was set as described in Section~\ref{sec:threshold}. The hyperparameters for the detection stage were determined using a grid search. Graph construction and traversal were handled using NetworkX~\cite{networkx}.

\subsection{Datasets}
We evaluate \methodName{} on three diverse datasets: the DARPA eng.3 THEIA and CADETS, and Public Arena datasets. These datasets provide a range of system behavior and attack scenarios on which to assess our method's effectiveness in detecting and tracing APTs.

\noindent\textbf{DARPA eng.3 THEIA:} The THEIA dataset~\cite{darpa} consists of 106 million system audit logs (40 GB uncompressed) collected from an Ubuntu Linux machine over 12 days. It includes a complex APT attack scenario carried out in two parts and a third unsuccessful attack.

\noindent\textbf{DARPA eng.3 CADETS:} The CADETS dataset~\cite{darpa} comprises 42 million system audit logs (25 GB uncompressed) collected from a FreeBSD system over 12 days. It provides granular system call traces and includes four APT attack scenarios, three of which are completely recorded and one of which is partially completed.

\noindent\textbf{Public Arena:} The Public Arena dataset~\cite{publicarena} comprises 16 million system audit logs (7 GB uncompressed) collected from two Windows hosts over six days, simulating a public cloud workspace environment. 

 From each dataset, we extract the \emph{data transfer} events, in order to capture the essential provenance information~\cite{watson,zengy2022shadewatcher}.

\begin{table}[htbp]
\centering
\caption{Train and test windows with detailed test provenance statistics}
\label{tab:train_test_provenance}
\resizebox{\textwidth}{!}{
\begin{tabular}{|c|c|cc|ccc|cc|}
\hline
\multirow{3}{*}{\textbf{Dataset}} & \multirow{3}{*}{\textbf{Split Time}} & \multicolumn{2}{c|}{\textbf{Train}} & \multicolumn{5}{c|}{\textbf{Test}} \\
\cline{3-9} 
 &  & Duration & \# Logs & Duration & \# Logs & \# Attacks & \multicolumn{2}{c|}{Provenance Statistics} \\
\cline{8-9}
 &  &  &  &  &  &  & Benign (N, E, KE) & Malicious (N, E, KE) \\
\hline
THEIA & 2018-04-09 22:15:12 & 3D13H & 9,654,772 & 7D11H & 17,827,942 & 2 & 492,556, 17,827,833, 1,823,963 & 16, 170, 16 \\
CADETS & 2018-04-06 11:00:00 & 6D8H & 4,239,474 & 3D4H & 10,949,668 & 3 & 263,775, 10,947,794, 1,293,534 & 33, 2,037, 54 \\
Public arena & 2022-05-13 00:00:00 & 3D9H & 2,593,769 & 7D9H & 6,468,573 & 1 & 25,527, 6,093,093, 106,285 & 6, 375,480,* 7 \\
\hline
\end{tabular}
}
\small\textit{N - nodes, E - edges, KE - key edges; *high because of repetitive communication between the malicious nodes}
\end{table}

\subsection{Train and Test Set Creation\label{train_test}}
\noindent \textbf{Train and Test Window.} For each dataset, we define a training window prior to the first attack scenario (see Table~\ref{tab:train_test_provenance} for the time windows used in each dataset). 
CADETS was trained on the first 28\% of the logs, THEIA was trained on the first 35\%, and Public Arena was trained on the first 29\%. This split mitigates data leakage and ensures that the evaluation reflects a real-world scenario with limited training data and prolonged inference periods. All methods are evaluated using the same train and test windows for each dataset.

\noindent \textbf{Ground Truth Establishment.\label{sec:groundtruth}}
We manually flag attack events in the raw logs based on the ground-truth documents provided with the datasets. 
These documents, which provide a thorough explanation of the attack scenarios, do not provide the specific logs associated with the attack. To identify key attack events, we rely on the assumption that data flow between two object entities in a provenance graph necessitates an intervening subject entity (i.e., a process). 
We begin by tagging file and socket nodes, using names from the ground-truth documentation. 
For each object entity identified, we locate its immediate neighboring subject entities and check for overlaps in their neighborhoods. 
Subject entities found in these overlaps are flagged as malicious processes due to their interactions with multiple attack-related object entities. 
Finally, we flag the provenance events between any two flagged entities as malicious, allowing us to reconstruct the attack story.

Then we map the attributes of each attack to the corresponding events in the raw logs. 
\emph{Key edges} are defined by merging duplicate events, distilling the core provenance trail and ensuring clarity and conciseness in the provenance representation.

Table~\ref{tab:train_test_provenance} presents the distilled statistics for each dataset, including the number of edges, nodes, malicious edges, malicious nodes, key edges, and dataset size.

\subsection{Evaluation Metrics and Compared Methods}

\noindent \textbf{Evaluation Metrics.}
We assess \methodName{}'s performance using a multi-layered evaluation framework that examines three levels of granularity: \emph{graph (attack) level}, \emph{node (entity) level}, and \emph{edge (event) level}. 
Each level evaluates different aspects of \methodName{}'s ability to detect and trace malicious activities in the system's provenance data. 
We use precision, recall, and the F1 score to measure detection accuracy and the false positive rate (FPR) to gauge alert relevance and \methodName{}'s ability to minimize alert fatigue. 
These metrics address the critical balance between detecting true threats and minimizing irrelevant alerts, which is essential for maintaining operational efficiency in real-world environments. 
Due to different graph partitioning techniques between methodologies, we do not measure the FPR at the graph level but note whether each method employs static or dynamic graph construction (DGC).

\noindent \textbf{Compared Methods.}
We compare \methodName{}'s performance to that of the following state-of-the-art APT detection methods: UNICORN~\cite{unicorn}, MAGIC \cite{jia2023magic}, DeepLog~\cite{du2017deeplog}, DISTDET~\cite{dong2023distdet}, and NODLINK~\cite{Li_2024}. These methods represent a diverse range of techniques, including graph-based approaches, tree-based-approaches, and sequence learning. 
We run UNICORN and MAGIC with the implementation and hyperparameters provided by the authors~\cite{han2020unicorn,jia2023magic} and DeepLog with the implementation and hyperparameters provided by DEEPCASE~\cite{van2022deepcase}; the results of NODLINK and DISTDET are sourced from the original papers.

\section{Results}
\subsection{Detection}

\begin{table}[htbp]
\centering
\centering
\caption{Graph level}
\label{tab:graph_level}
\resizebox{0.6\textwidth}{!}{%
\begin{tabular}{|c|c|cccc|}
\hline
\textbf{Dataset} & \textbf{Method} & R & P & F1 & DGC \\ \hline
\multirow{3}{*}{THEIA} & \methodName{} & 1.00 & 1.00 & 1.00 & \checkmark \\
 & UNICORN & 1.00 & 1.00 & 1.00 & - \\
 & DISTDET* & 1.0 & 0.98 & 0.98 & \checkmark  \\ \hline
\multirow{3}{*}{CADETS} & \methodName{} & 1.00 & 1.00 & 1.00 & \checkmark \\
 & UNICORN & 1.00 & 0.98 & 0.98 & - \\
 & DISTDET* & 1.0 & 0.98 & 0.98  & \checkmark \\ \hline
\multirow{2}{*}{Public Arena} & \methodName{} & 1.00 & 1.00 & 1.00 & \checkmark \\
 & DISTDET* & 0.9 & 0.88 & 0.89 & \checkmark \\ \hline
\end{tabular}
}

\small \textit{R - Recall, P - Precision, F1 - F1 Score, FPR - False Positive Rate, DGC - Dynamic Graph Construction;}
\small \textit{*NODLINK and DISTDET results from original papers}
\end{table}
\begin{table}[htbp]
\centering
\caption{Node level}
\label{tab:node_level}
\resizebox{0.6\textwidth}{!}{%
\begin{tabular}{|c|c|cccc|}
\hline
\textbf{Dataset} & \textbf{Method} & R & P & F1 & FPR \\ \hline
\multirow{4}{*}{THEIA} & \methodName{} & 1.00 & 0.55 & 0.71 & 0.00 \\
 & UNICORN & 1.00 & 0.00 & 0.00 & 0.23 \\
 & NODLINK* & 1.00 & 0.23 & 0.37 & 0.00 \\
 & MAGIC & 1.00 & 0.32 & 0.48 & 0.00 \\ \hline
\multirow{4}{*}{CADETS} & \methodName{} & 1.00 & 0.74 & 0.85 & 0.00 \\
 & UNICORN & 1.00 & 0.00 & 0.00 & 0.27 \\
 & MAGIC & 1.00 & 0.09 & 0.17 & 0.04 \\
 & NODLINK* & 1.00 & 0.14 & 0.25 & 0.01 \\ \hline
\multirow{1}{*}{Public Arena} & \methodName{} & 1.00 & 0.67 & 0.80 & 0.00 \\\hline
\end{tabular}
}
\end{table}
\begin{table}[htbp]
\centering
\caption{Edge level}
\label{tab:edge_level}
\resizebox{0.6\textwidth}{!}{%
\begin{tabular}{|c|c|cccc|}
\hline
\textbf{Dataset} & \textbf{Method} & R & P & F1 & FPR \\ \hline
\multirow{2}{*}{THEIA} & \methodName{} & 1.00 & 0.42 & 0.59 & 0.00 \\
 & Deeplog & 0.89 & 0.00 & 0.00 & 0.04 \\ \hline
\multirow{2}{*}{CADETS} & \methodName{} & 0.91 & 0.35 & 0.51 & 0.00 \\
 & Deeplog & 0.86 & 0.00 & 0.00 & 0.04 \\ \hline
\multirow{2}{*}{Public Arena} & \methodName{} & 1.00 & 0.44 & 0.61 & 0.00 \\
 & Deeplog & 1.00 & 0.00 & 0.01 & 0.01 \\ \hline
\end{tabular}
}

\end{table}

\noindent \textbf{Graph (Attack) Level Detection.}
Table~\ref{tab:graph_level} presents the results at the graph level, where \methodName{} achieves perfect recall and precision, outperforming DISTDET~\cite{dong2023distdet} and matching the performance of UNICORN~\cite{han2020unicorn}. However, UNICORN generates static graphs containing 100,000 nodes, providing alerts at a coarse granularity and requiring significant effort from security analysts to pinpoint malicious activity. 

\noindent \textbf{Node (Entity) Level Detection.}
Table~\ref{tab:node_level} presents the results at the node level, where \methodName{}, MAGIC, and NODLINK achieve perfect recall on the CADETS dataset, successfully identifying all malicious entities. However, \methodName{} outperforms all other methods in terms of precision across all datasets while maintaining a consistently low FPR. UNICORN, at a finer granularity, fails to provide precise alerts. NODLINK utilizes 80\% of the data for training and only 20\% for testing, while \methodName{} achieves better performance using only 30\% for training and 70\% for testing. Using the same split, MAGIC struggles with concept drift; this demonstrates \methodName{}'s ability to effectively detect attacks with limited training data, making it more practical for real-world scenarios.

\noindent \textbf{Edge (Event) Level Detection.}
Table~\ref{tab:edge_level} presents the results at the edge level, where \methodName{} consistently outperforms DeepLog in terms of recall, precision, and F1 score while maintaining a lower FPR across all datasets. This highlights \methodName{}'s superiority in accurately identifying malicious events and paths in the system's provenance data. UNICORN, MAGIC, NODLINK, and DISTDET do not provide effective results at the edge level, limiting their ability to capture the full attack story.

In summary, \methodName{} demonstrates superior APT attack detection performance compared to state-of-the-art methods across various granularity levels. Its ability to generate concise and informative alerts, handle concept drift, and adapt to different system environments makes it a promising solution for real-world APT detection and investigation. \methodName{}'s focus on generating high-quality and precise alerts, along with its adaptive alert generation and reduction of noise in attack graphs, makes it an effective tool for in-depth APT investigation and forensic analysis.
 
\subsection{Alert Quality}
System provenance data provides the information needed to capture crucial interactions between entities within a system. 
\methodName{} leverages this data to generate alerts that highlight anomalous activities along key paths. 
High-quality alerts, with detailed attack narratives, are essential for security analysts to quickly and decisively take action. 
In this section, we analyze the quality of alerts generated by \methodName{} by examining examples from the CADETS, THEIA, and Public Arena datasets, and present the performance metrics across all the attacks.

\begin{figure}[htbp]
\centering
\hfill
\includegraphics[width=1.0\textwidth]{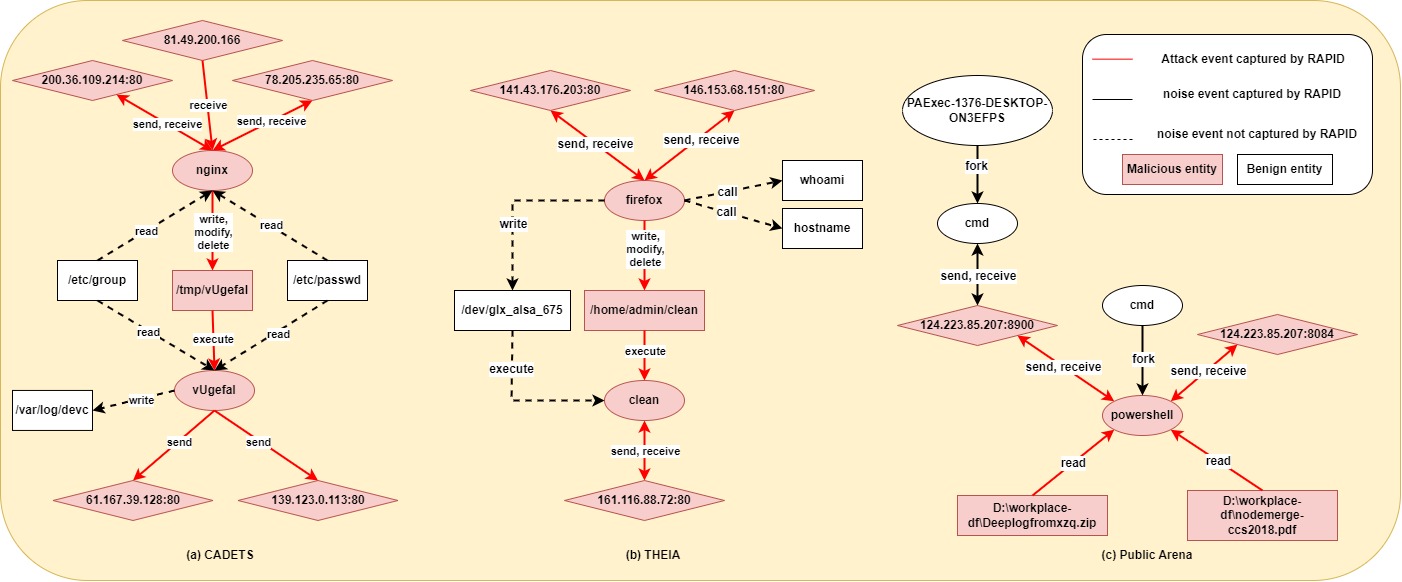}
\caption{Alert graphs generated by \methodName{} for attacks in the CADETS (left), THEIA (middle), and Public Arena (right) datasets.}
\label{fig:alert_graph_examples}
\end{figure}

Figure \ref{fig:alert_graph_examples} shows alert graphs generated by \methodName{} for APT attacks in the CADETS, THEIA, and Public Arena datasets. 
Each graph provides a narrative of the attack, showcasing how \methodName{} identifies and correlates malicious activities.

For instance, the alert from the CADETS dataset captures the following flow of malicious events: external networks interact with \textit{nginx}, which then elevated the permissions and executed a malicious binary \textit{VUgefal}. 
\textit{VUgefal} then attempted to inject data into \textit{/var/log/devc} and finally communicates with external networks.  
This narrative enables a security analyst to understand the alert, quickly respond, and filter out false alarms. 
The clarity and granularity in the alert can significantly reduce investigation time and improve the accuracy of incident response.

The usefulness of these alerts extends to other datasets as well. In the THEIA and Public Arena datasets, \methodName{} effectively identifies the primary steps in the attack process, demonstrating its robustness and versatility in handling different types of attacks.

\begin{table}[htbp]
\centering
\caption{Edge level performance per attack}
\label{tab:alert_metrics}
\begin{tabular}{|c|cccccc|}
\hline
\textbf{Dataset} & \textbf{Instance} & \textbf{R} & \textbf{P} & \textbf{TP} & \textbf{FP} & \textbf{Score} \\ \hline
\multirow{2}{*}{THEIA} & Attack 1 & 1.00 & 0.42 & 16 & 22 & 0.94 \\
                    & Attack 2 & 1.00 & 0.42 & 16\footnote{Overlaps with first attack} & 22 & 0.94 \\ \hline
\multirow{3}{*}{CADETS} & Attack 1 & 0.42 & 0.36 & 17 & 30 & 0.83 \\
 & Attack 2 & 1.00 & 0.54 & 13 & 11 & 0.97 \\
 & Attack 3 & 1.00 & 0.43 & 19 & 25 & 0.91 \\ \hline
\multirow{1}{*}{Public Arena} & Attack 1 & 1.00 & 0.64 & 7 & 4 & 0.84 \\ \hline
\end{tabular}

\small \emph{TP - \# Key events, FP - \# Benign events}
\end{table}

Table \ref{tab:alert_metrics} presents the edge-level performance metrics for each attack. 
We specify the specific number of key events (TP) and benign events (FP) presented in the alert graphs to better quantify the graph size and quality.
The high recall values indicate that \methodName{} successfully identifies the key events of the attacks, while the precision values show its effectiveness in filtering out benign activities. 
\methodName{}'s consistent performance across different datasets underscores its robustness and practicality for real-world applications.

\subsection{Impact of Object Embedding\label{sec:impact_embeddings}}
\methodName{} leverages object embeddings in a two-phase approach aimed at enhancing both anomaly detection and alert tracing. 
In this section, we aim to evaluate the impact of this approach on the performance of \methodName{} by conducting two ablation studies. 
In the first ablation study, we focus on the effect of the object embeddings on the detection phase, while in the second we focus on the impact of the object embeddings during the tracing phase.

\noindent \textbf{Anomaly Detection Accuracy.}
The object embeddings enable \methodName{} to capture nuanced relationships between system entities, resulting in more accurate pattern identification during the detection phase. An ablation study comparing the performance of \methodName{} with and without using the embeddings as enriched input for the anomaly detector shows a significant improvement in precision from 0.29 to 0.67 when embeddings are used, while maintaining a recall of 1.0 in both cases. This highlights the embeddings' ability to differentiate between benign and malicious activities more effectively.

\begin{figure}[htbp]
    \centering
    \includegraphics[width=1.0\textwidth]{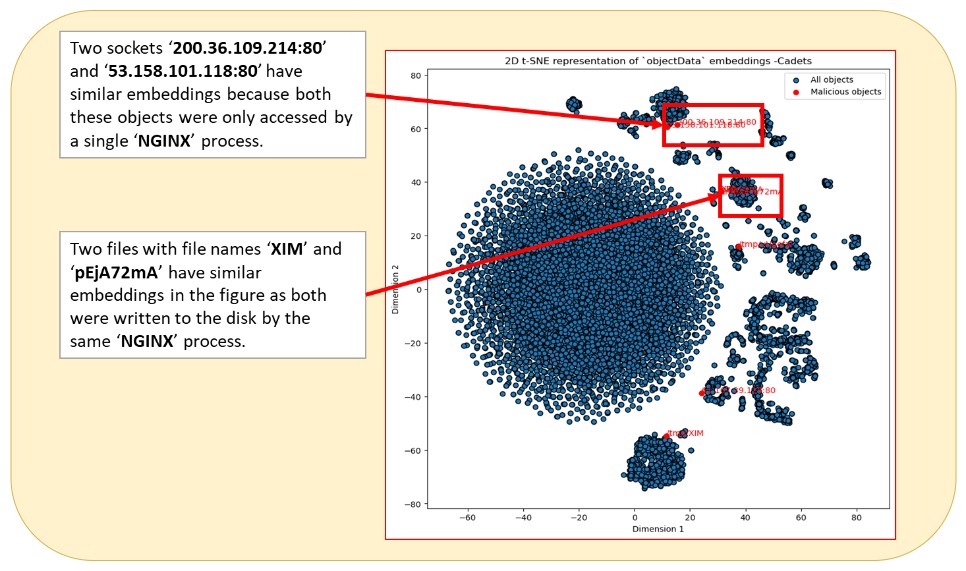}
    \caption{Illustration of the latent representation of the object embeddings using T-SNE for dimension reduction.}
    \label{fig:embedding_explanation}
\end{figure}

\noindent \textbf{Clustering and Provenance Context.}
The object embeddings facilitate the effective clustering of system entities based on behavioral similarities. Figure \ref{fig:embedding_explanation} illustrates how objects interacting with the same process tend to cluster together in the embedding space. 
While attacks cannot be detected using solely clustering, it prioritizes entities for correlation during alert tracing~(Section \ref{sec:alert_generation}), guiding the anomaly-to-attack graph propagation towards the most suspect behaviors and refining the sub-graph.
We conduct an additional ablation study on the object embeddings during the alert tracing phase. Our results show that embedding-based clustering significantly reduces the attack graph size from 1,746.5 to 53.8 edges on average and reduces redundancy from 96\% to 44\%, highlighting the embeddings' ability to capture the most relevant attack chain components and generate more concise and actionable alerts.

In conclusion, the two-phase leveraging of object embeddings in \methodName{} significantly enhances anomaly detection accuracy and enables precise alert tracing, thus reducing graph size and improving results. The context captured by the embeddings provides comprehensive understanding of system behavior, facilitating the identification of malicious activities. This impact makes the object embeddings' two-phase integration a valuable component of \methodName{}'s novel approach to APT detection and tracing.

\subsection{Scalability and Performance Overhead}
Efficient resource utilization and managing scalability are critical for the practical deployment of APT detection systems in real-world scenarios. \methodName{} leverages lightweight neural network models, LSTM and CBOW~\ref{sec:method}, specifically chosen for their efficiency and scalability.

In our performance evaluation on the THEIA dataset, \methodName{} achieved a throughput of 3.6×$10^4$ logs per second during inference, significantly outperforming MAGIC's 1.0×$10^4$ logs per second and matching UNICORN's 3.4×$10^4$ logs per second. This throughput suggests that \methodName{} can effectively monitor over 1,300 hosts on a single server, highlighting its ability to handle vast amounts of log data which is crucial for timely threat detection and response in large-scale enterprise networks.

\methodName{}'s design and empirical results illustrate its ability to balance high performance with reduced computational cost, making it a scalable solution for monitoring extensive network infrastructures.

\section{Conclusion\label{sec:conclusion}}
In this paper, we introduce RAPID, a novel approach for APT detection and tracing that synergistically integrates self-supervised anomaly detection, provenance graph analysis, and unsupervised learning techniques. 
RAPID's key innovation lies in its dual-purpose application of dynamically updated object embeddings, which enhances both anomaly detection accuracy and alert tracing capabilities within the provenance graph.

By capturing the nuanced relationships between system entities, \methodName{}'s context-aware anomaly detection significantly improves the identification of malicious activities.
Leveraging object embeddings in the alert tracing phase enables \methodName{} to generate high-granularity alerts that accurately portray the attack kill chain, bridging the gap between detection and investigation, reducing alert fatigue, and strengthening an organization's security posture against advanced persistent threats.

\methodName{} outperforms state-of-the-art methods across multiple levels of granularity, achieving near-perfect precision and recall while minimizing alert noise. The method's runtime and space complexity analysis highlight its practicality for real-world deployments, showcasing its ability to handle large-scale, complex systems efficiently.

\noindent \textbf{Limitations and Future Work.} 
Despite the robustness of \methodName{} in both capturing attacks and filtering alerts, it still requires periodic updates to avoid concept drift. Future work can explore automatically integrating benign data back into the model to concurrently learn and refit itself. This would enhance \methodName{}'s adaptability and ensure its effectiveness in dynamic environments.

Although \methodName{}'s design choice stems from a focus on scalability and practicality, more complex node embeddings could further enhance the model. However, this must be done with careful consideration of time and space costs to maintain the balance between performance gains and computational efficiency.

The primary next step forward, to further build on \methodName{}, would be to integrate it into an automated threat response system. The high granularity and effectiveness of the alerts will provide valuable input to automated threat response pipelines, enabling swift and targeted remediation actions.

\bibliographystyle{splncs04}
\bibliography{main}

\appendix

\section{Alert Ranking\label{sec:alert_ranking}} 
Building upon the alert tracing approach described in Section \ref{sec:alert_generation}, \methodName{} introduces a novel alert ranking mechanism that maximizes the use of information gained from both the anomaly detector and the provenance graph. The ranking process integrates two key graph features: anomaly scores of each edge and kill chain components, enabling security analysts to prioritize alerts effectively.

\noindent \textbf{Kill Chain Components.}
To provide additional context and depth to the alert ranking process, \methodName{} leverages the MITRE ATT\&CK framework~\cite{strom2018mitre}. The system extracts Tactics, Techniques, and Procedures (TTPs) from the attack graph, identifying patterns and behaviors indicative of different stages of the cyber kill chain. This approach is inspired by previous works in the field, such as ATLAS~\cite{alsaheel2021atlas}, NEEDLE~\cite{yu2019needle}, and Krystal~\cite{kurniawan2022krystal}. While these tags do not directly influence the detection pipeline, they enrich the analysis of generated alerts with valuable context.

\noindent \textbf{Alert Ranking Metric.}
\methodName{} introduces a scoring system that blends the anomaly scores detailed in Section~\ref{sec:threshold} and kill chain components to rank alerts. The metric is defined as follow
\[
\textit{Attack Score} = \frac{1}{2N} \sum_{i=1}^{N} \textit{Anomaly Score}_i + \frac{1}{2} \left( \frac{K_{achieved}}{K_{total}} \right)
\]
Where:
\begin{itemize}
\item $N$ is the total number of events in the attack graph.
\item \textit{$Anomaly Score_i$} refers to the anomaly score of each event, assigned by the anomaly detector.
\item $K{achieved}$ represents the number of unique Kill Chain Phases achieved, based on the extracted MITRE ATT\&CK TTPs.
\item $K_{total}$ is the total number of Kill Chain Phases outlined by the MITRE ATT\&CK framework.
\end{itemize}

This metric ensures that attack graphs with a higher concentration of anomalous events and a wider coverage of kill chain phases are assigned higher scores. The scores are normalized to a range between 0 and 1, facilitating easy comparison and prioritization of alerts.

\methodName{} presents the ranked alerts in a graph format, prioritizing them by significance and urgency. Each alert includes the attack score, the associated MITRE ATT\&CK techniques, and the anomalous events with their relationships within the attack graph. This comprehensive and actionable format empowers security analysts to efficiently triage, investigate, and respond to the most critical threats.

\noindent \textbf{Alert Presentation.}
\methodName{} presents the ranked alerts in a graph format, prioritizing them based on their attack scores. Each alert includes The attack graph, attack score and the associated MITRE ATT\&CK TTPs, 
By integrating anomaly scores and kill chain components, \methodName{}'s alert ranking approach provides a powerful and context-rich mechanism for prioritizing security threats. This hybrid approach leverages the strengths of both unsupervised learning and domain knowledge, enabling security analysts to focus their efforts on the most significant and impactful alerts.

\end{document}